# Green IoT using UAVs in B5G Networks: A Review of Applications and Strategies

S. H. Alsamhi[1,2], Fatemeh Afghah[3], Radhya Sahal[4], Ammar Hawbani[5], A. A. Al-qaness[6], B. Lee[1], Mohsen Guizani[7]

[1]Software Research Institute, Athlone Institute of Technology, Athlone, Ireland (e-mail: salsamhi@ait.ie).
[2]IBB University, Ibb, Yemen.
[3]School of Informatics, Computing and Cyber Systems, Northern Arizona University, USA
[4]Hudaidah University, Hudaida, Yemen.
[5]School of Computer Science and Technology; University of Science and Technology of China, Hefei, China.
[6] State Key Laboratory for Information Engineering in Surveying, Mapping and Remote Sensing, Wuhan University, China.
[7]Department of Computer Science and Engineering, Qatar University, Qatar

*Abstract*: Recently, Unmanned Aerial Vehicles (UAVs) present a promising advanced technology that can enhance people life quality and smartness of cities dramatically and increase overall economic efficiency. UAVs have attained a significant interest in supporting many applications such as surveillance, agriculture, communication, transportation, pollution monitoring, disaster management, public safety, healthcare, and environmental preservation. Industry 4.0 applications are conceived of intelligent things that can automatically and collaboratively improve beyond 5G (B5G). Therefore, the Internet of Things (IoT) is required to ensure collaboration between the vast multitude of things efficiently anywhere in real-world applications that are monitored in real-time. However, many IoT devices consume a significant amount of energy when transmitting the collected data from surrounding environments. Due to a drone's capability to fly closer to IoT, UAV technology plays a vital role in greening IoT by transmitting collected data to achieve a sustainable, reliable, eco-friendly Industry 4.0. This survey presents an overview of the techniques and strategies proposed recently to achieve green IoT using UAVs infrastructure for a reliable and sustainable smart world. This survey is different from other attempts in terms of concept, focus, and discussion. Finally, various use cases, challenges, and opportunities regarding green IoT using UAVs are presented.

## I. Introduction

Recently, UAV technology has emerged as a promising technology for intelligent surveillance and reconnaissance operations. The use of UAVs leads to enormous benefits such as reduced transmission power consumption [1], conservation of resources [2], and reduced pollution [2, 3]. Furthermore, UAVs prove themselves indispensable in natural and human-made disaster relief operations (in preparedness, assessment, and recovery), emergency aid, parcel delivery, imagery over large geographical areas, data gathering, and environmental monitoring [4-6]. Recently, the UAVs technology has been classified as a promising solution for combatting COVID-19 due to the capability of UAVs in detecting a person when not wearing a mask (when a UAV is equipped with a high-resolution camera), identifying an infected person (when a UAV is equipped with a thermal camera) [7]. All of the above characteristics lead to improved smartness and reliability of the smart environment. The UAV can act as an aerial Base Station (BS) for delivering communication services to a wide coverage area. Agility and the ability to enable communication beyond Line of Sight (LoS) in disaster areas are among the most important drones' features, making UAVs play a vital role in achieving the green Internet of Things (IoT). In areas not accessible by human beings, such as disaster areas, the UAVs can act as aerial relays where they are equipped with IoT devices and advanced communication equipment to deliver communication services to survivors and share collected data for decision making and taking necessary actions in real-time [8]. For instance, the UAVs can guide Search and Rescue (SAR) teams when their wearable smart devices are connected to a UAV to direct them towards a suitable route for their safety [9].

IoT devices are not able to send data over considerable large distances because of their energy limits. Therefore, a UAV can represent a crucial technology in IoT communication, and data aggregation from small devices, such as sensors and health care devices [10]. In extending the range of communication over IoT devices, the integration of UAVs and IoT is required to gather data from IoT devices and transmit it to the intended destination [11]. Authors of [12] provided a solution to integrate UAVs and IoT devices into disaster response settings and offered event detecting and automatic network repair facilities. An advantage of this technology is that the UAV can deliver IoT data in real-time to any particular place or the ground control for agile decision-making.

For ensuring energy efficiency on these low-power devices, smart UAVs can help by moving automatically towards potential smart IoT devices, establishing the connection, collecting data, and transmitting onward (to the edge/fog/cloud), which is out of the transmission range of smart IoT devices [13]. Fig.1 shows the UAVs capability as a Fog node to collect data from many IoT devices in a large area and control the operation mode (sleep or active) of distributed devices for saving energy. The devices shown in red are in active mode, while those shown in black are in sleep mode. Thus, a smart UAV can also help in ensuring battery longevity.

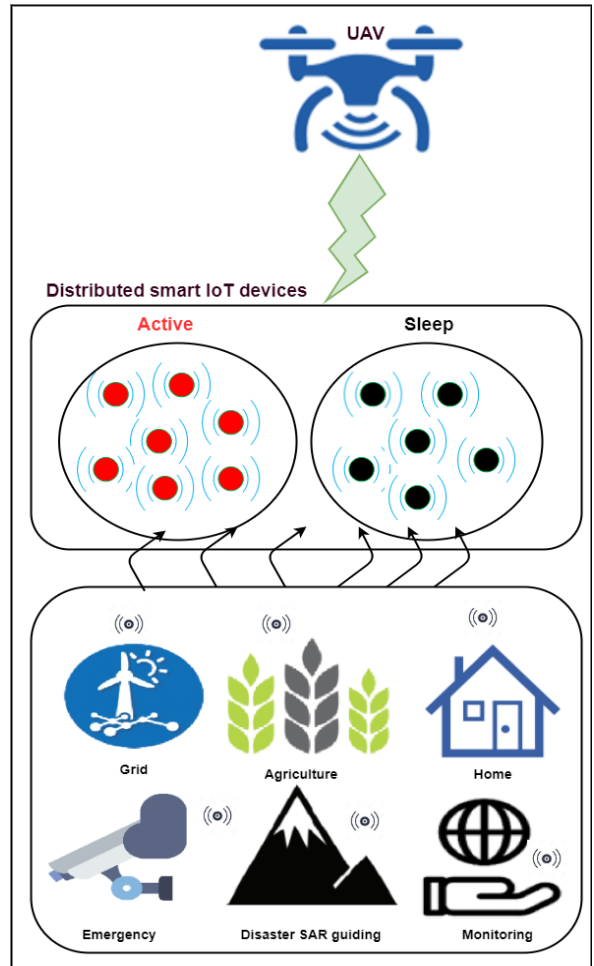

Fig.1 UAV gather data from IoT devices

Furthermore, a UAV can be used to connect to IoT devices effectively and efficiently to deliver broadband communication services to large coverage areas at low cost in order to decrease energy consumption [14], and reduce transmission power while retaining reliability [1]. Recently, connected swarm UAVs with IoT devices have also been considered [15, 16]. In [17], the authors presented a steering technique for UAVs with multiple 4G networks to guarantee reliable connectivity that supports Quality of Service (QoS) with an acceptable energy consumption. One of the primary applications of the 5G communication network is IoT and sensor networks. Qualcomm and Nokia have initiated drone projects, in which they conducted many experiments on how to use UAVs for Beyond 5G (B5G) communication networks [18, 19]. Many surveys and tutorials analyze UAVs' challenges, opportunities, and future for B5G [20-23]. Collaboration between IoT and UAVs can be employed to achieve new IoT regimes, such as civil applications, industrial, and medical environments, and extend the UAV functionalities [24]. Therefore, in [25], authors considered UAVs as the middleware for energy-efficient discovery in sensor networks and IoT.

Nowadays, IoT has enabled advanced applications in smart homes, intelligent logistics, smart cities, smart power grids, smart manufacturing, etc. These applications have become critical to our lifestyle. The environmental factor pertains to the use of green things that protect environment from hazardous and harmful emissions, alleviate the consequences of climate change, conserve natural resource usage, and reduce pollutions and power consumption. Hence, greening IoT networks is a critical technology towards reducing $CO_2$ emissions and pollution, environmental conservation, and minimization of operating costs and power consumption [26]. Furthermore, utilizing UAVs as a bridge (fog computing layer) for monitoring Industrial IoT (IIoT) in smart industries [27, 28], has gained considerable attention. UAV-enabled Mobile Edge Computing (MEC) is empowered to support offloading tasks and improve computing capability and energy efficiency [29]. To the best of our knowledge, existing research attempts have not addressed UAV technology's importance to green IoT applications to improve smart cities' quality of life. Therefore, this paper provides a comprehensive survey of strategies and techniques used for greening IoT applications using UAVs to facilitate energy efficiency, sustainable QoS, and delivery of services to large coverage areas at low cost, thereby reducing the carbon footprint.

**A. Contributions**

This paper is designed to provide a comprehensive overview of the techniques and strategies used for greening IoT using UAV technology. To this end, the primary goal is to collect the literature on the use of UAVs for greening IoT that address the applications, challenges, and opportunities of UAVs as edge intelligence devices for greening IoT. We believe that this comprehensive survey will help readers to fully accomplish the potential strategies for greening IoT using UAVs edge intelligence technology. Furthermore, we will highlight the applications of green IoT by using UAVs technology. The contributions of this paper are listed as follows:

- **Green IoT**, we outline the critical role of greening IoT by integrating UAVs with IoT. We discuss the strategies and techniques used for UAVs use cases for green IoT applications based on enhancing green IoT such as smart homes, smart cities, smart agriculture, smart healthcare, and smart grid.
- **Connectivity and communication beyond 5G**, we discuss the most current development regarding integrating UAVs and B5G networks. We introduce UAVs' important applications for enabling Information and Communication Technologies (ICT) such as sensors, 5G and beyond 5G technologies, and Radio Frequency IDentification (RFID) to green IoT.
- **Energy efficiency**, we discuss the most recent studies regarding the UAV technology's performance in improving IoT energy efficiency. We focus on strategies and techniques for power optimization problems of integration of UAVs and IoT. Furthermore, we summarize various efficient techniques for energy efficiency of UAV systems to green IoT.
- **Pollution hazardous monitoring**, we discuss the most recent studies in regards to monitoring pollution using UAVs. We discuss the strategies and techniques that enable UAVs to monitor and improve air quality effectively and efficiently. Then, we summarize various efficient techniques for reducing pollution with the help of UAVs in Industry 4.0 applications.
- We discuss future scopes, challenges, and opportunities of using UAVs technology for greening IoT.

**B. Review of recent surveys**

The IoT regime includes the ever-expanding world of sensors, computers, actuators, and smart devices. For example, everything could communicate over the Internet, such as buildings, smartphones, vehicles, home appliances, and even natural objects, to create a smart world. However, many IoT devices require high energy to cooperate, connect, and share information with the central node (if applicable) or one another. Therefore, developing effective power conservation strategies becomes critical for extending the battery lifetime. Furthermore, load balancing can play a critical role in improving network lifetime by decreasing smart devices' energy consumption. Therefore, many studies have surveyed appropriate strategies and technologies to improve the intelligent world's smartness level by applying UAVs for different applications [30-33]. Smartness is evaluated by many factors such as the smart economy, smart living, smart mobility, smart environment, smart people, smart governance, smart tourism, etc. In [31], the authors introduced technical problems and potential applications related to smart UAVs in smart cities. The authors discussed enabling techniques used to develop the integration of smart UAVs in smart cities. However, they did not address enabling technologies for green IoT by using UAVs technology in smart city applications. Also, in [32], the authors summarized the strategies and techniques that are used for greening IoT and improved the smartness of cities. However, all of them addressed technologies that did not discuss how UAV technology can improve green IoT applications.

The authors in [34] described different frameworks for smart city monitoring by using smart drones. Motlagh *et al.* [35] reviewed how UAVs operations support IoT services. The authors discussed the possibility of UAVs being equipped with IoT devices such as sensors, RFID devices, and cameras to collect data from monitoring areas. Also, in [33], the authors described drones' concepts equipped with IoT devices being used to deliver services to a wide coverage area. This paper mainly discussed how UAVs could be utilized to support crowd surveillance, depending on face recognition techniques. They compared the processing of video data on MEC and UAV payload. The findings showed the MEC-based offloading technique's superiority, as demonstrated by lesser processing time and energy consumption. Hence, gathering data from IoT devices on the ground with improving energy efficiency was not addressed. Loke [36] outlined the challenges and open issues related to fog computing and UAVs for efficient data collection.

However, in [30], the authors extended the collaboration between UAVs and IoT for enabling connectivity. The finding showed that the connectivity rate was reduced with increased UAVs' height because more significant interference occurred by increasing the path loss. In [37], the authors discussed the opportunities, challenges, and applications of UAVs in smart cities. Moreover, cybersecurity issues, public safety, and privacy issues related to the application of UAVs for smart cities were discussed in [38]. Many researchers have discussed green IoT with the help of UAVs, but for specific applications and purposes such as security [39, 40], energy efficiency [41] networking, 5G [42, 43], and autonomy [44]. Furthermore, the QoS from space technologies is considered to improve greening communication [45-48], public safety [49], and disaster recovery [50, 51]. The authors in [52] discussed the use of using UAVs for green logistics, healthcare, and agriculture. The authors in [41] studied a UAVs-assisted wireless system and focused on maximizing the vehicle's rate on the ground to join the UAV's trajectory while maintaining UAV's energy constraints. Furthermore, the optimal trajectory is introduced for balancing QoS and energy efficiency intelligently using Q-learning [53-56], RF band allocation [57], and wireless power transfer to recharge the UAVs [58].

Briefly, the surveys in [35] and [33] are primarily limited to UAVs' architecture to deliver IoT services. Also, [59] introduced the benefits and applications of UAVs for wireless telecommunications. More so, [31] and [29] discussed enabling technologies that integrate smart cities and UAVs to be greener and smarter. Additionally, in [38], the authors only addressed UAVs for public safety and privacy, monitoring smart cities [34]. Surprisingly, no survey has been thoroughly carried out, given the importance of using UAVs to green IoT and enhance the smart world's quality of life. In light of the above, Table1 shows a comparison of the related work in this area.

Table.1 Comparison of existing surveys

| Ref. | Highlighted | UAV | IoT | UAV and IoT | 5G | Energy Efficiency | ICT | Pollution |
|---|---|---|---|---|---|---|---|---|
| [31] (2020) | Integrated UAVs in smart cities. | √ | | | | | | |
| [35] (2016) | UAV architecture for delivering IoT services. | √ | √ | | | | | |
| [32] (2019) | Greening IoT for smart cities applications | | √ | | | √ | √ | √ |
| [34] (2018) | Smart UAV for smart city monitoring | √ | | | | | | |
| [33] (2017) | Drone-based IoT platform for crowd surveillance. | √ | √ | | √ | | | |
| [52] (2020) | UAV technology as a green tactical solution for logistics, healthcare, and agriculture. | √ | | | | √ | | √ |
| [60] (2015) | Green IoT and smart world | | √ | | √ | √ | √ | √ |
| [42] (2019) | Multi-UAV for network management and deployment to achieve energy efficiency | √ | | | | √ | | |
| [39] (2020) | UAV's security issues for green IoT agriculture | √ | √ | √ | | | | |
| [41] (2020) | UAV to ground integrated green IoT | √ | √ | √ | | √ | | |
| [43] (2020) | UAV-assisted B5G applications | √ | √ | | √ | √ | | |
| [44] (2021) | Autonomy of UAV for ecosystem applications | √ | √ | | | √ | | |
| This work | Potential techniques and strategies proposed recently to achieve green IoT by using UAV for reliable and sustainable smart cities. | √ | √ | √ | √ | √ | √ | √ |

## C. Study scope and survey structure

UAV technology plays a vital role in improving connection accuracy, transmitting and gathering IoT data by improving the metrics such as transmission energy efficiency, QoS, data collection efficiency, and reducing pollution hazards of ICT. This paper presents a comprehensive survey on efficient and intelligent UAV techniques, which have been proposed in the literature in recent years for greening IoT. The paper focuses on different criteria for continuing connectivity between UAVs and IoT devices to improve QoS metrics (path loss, Received Signal Strength (RSS), bandwidth, drop rate delay), and energy efficiency and reducing hazardous pollution.

The rest of the survey is organized as shown in Fig.2. The overview of using UAV technology for green IoT is discussed in Section II. Section III, IV, and V discuss how the UAVs are used for energy efficiency, green ICT, and reducing pollution hazards. Applications of UAVs are discussed in Section VI. Then, the future scope and conclusion are discussed in Section VII and VIII, respectively.

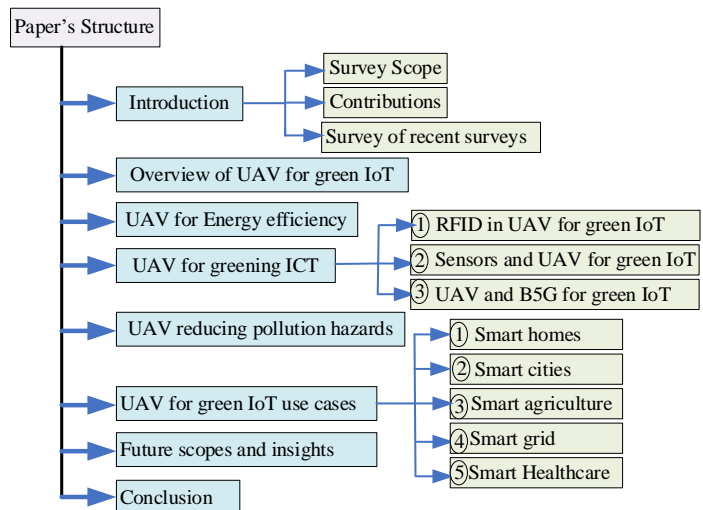

Fig.2 survey structure

## II. Overview of using UAVs for greening IoT

To green IoT, energy efficiency plays a crucial role in achieving the reliability and sustainability of services in smart cities [61]. In networking, green IoT aims to determine the relay location and the number of nodes that satisfy the power consumption and budget constraints. Hence, the devices are proposed to be equipped with sensors and communication add-ons. They can sense the surrounding things and communicate with each other. Green IoT focuses on green design, green manufacturing, green utilization, and green disposal [62]. Arshad *et al.* [63] evaluated efficient techniques for greening IoT based on reducing energy consumption.

Green IoT includes design and leverage aspects, as shown in Fig. 3. The design aspect includes communications protocols, designing energy-efficient computing devices, and networking architectures for interconnecting physical applications in smart cities [60]. Leveraged technologies are used for improving IoT in terms of cutting pollution hazards and enhancing energy efficiency. Moreover, green IoT involves various strategies and techniques for reducing energy consumption, reducing CO2 emissions, and reducing IoT devices' pollution. UAVs play an essential role in enhancing IoT devices' performance to improve life quality and green environment.

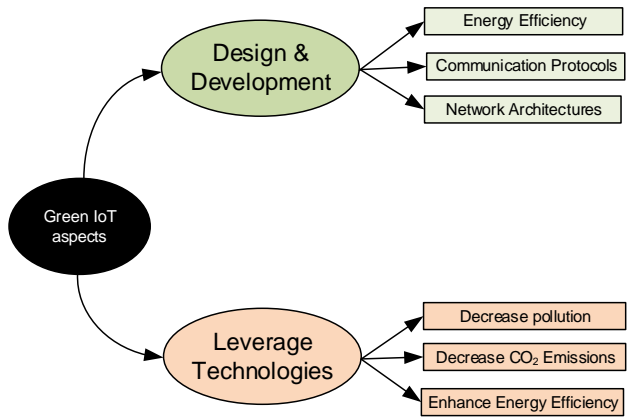

Fig.3 Green IoT design and leverage aspects

Recently, UAVs represent a microcosm of IoT. The advantages of using a UAV for green IoT include better connectivity and lower prices. Equipment such as sensors and cameras are used for remote control and navigation of the drones. UAVs can collect data from the ground of any environment by IoT devices carried in the UAV payload. The drones' service delivery is implemented using wireless communication technologies, including 5G, LTE, and Wi-Fi.

Connecting many UAVs (drone swarms) in the sky will enable many new services and business opportunities. Noting UAVs' mobility and agility, swarms can enhance the QoS and improve temporary events' capacity and coverage in 5G cellular networks [64]. Fig. 4 shows how UAVs can deliver broadband services to large coverage areas. For providing smart city monitoring, disaster recovery, and traffic monitoring, multi-UAV swarms can be used effectively. For instance, in [65], the UAV was used for delivering communication services when conventional wireless networks are destroyed, for supporting disaster recovery and SAR operations [66]. Similarly, Merwaday *et al.* [67] addressed how to use UAVs as a Base Station (BS) during a disaster for public safety and guiding SAR teams to perform complex tasks efficiently. In addition to the work of [65, 67], Bor-Yaliniz *et al.* [68] envisioned UAV-assisted cellular networks for delivering wireless communication networks anywhere as per demand at any time. Predicting the coverage area of the UAV is of utmost importance for successfully performing the intended application. Alsamhi *et al.* [69] applied an intelligent technique for predicting the signal strength and demonstrated that the coverage area could be identified based on the signal strength. The authors in [70] introduced a solution for cooperative spectrum leasing among UAVs and cellular users during the disaster situations provide additional spectrum access for the UAVs in return to enhancing the quality of communications for cellular users through cooperative relaying.

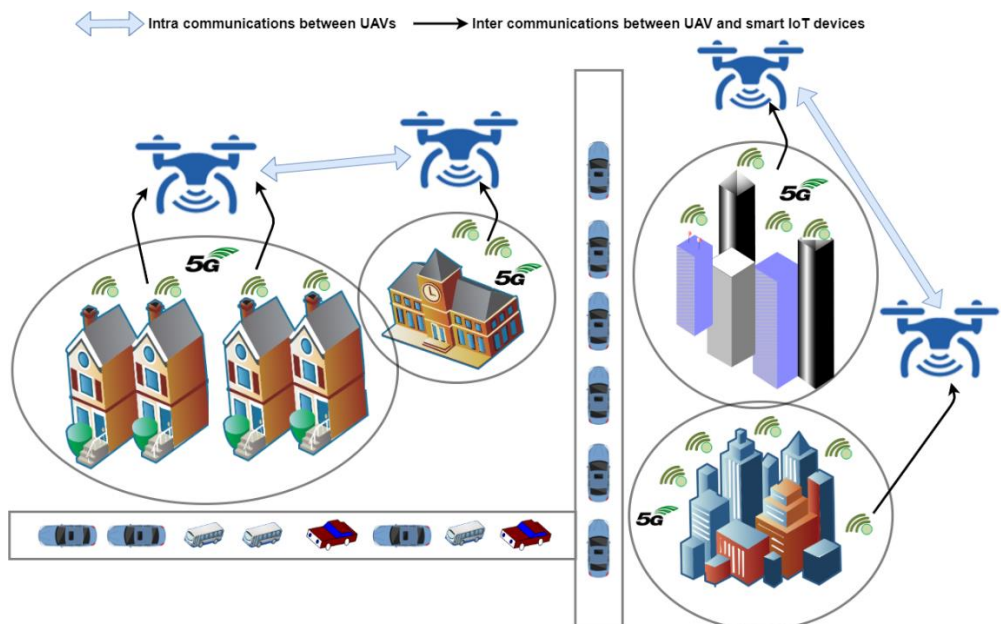

Fig.4 Multi-UAV for providing services to large coverage areas

Furthermore, a statistical method to predict the path loss between a UAV and a terrestrial wireless terminal was presented in [71]. The authors of [72] also addressed the path loss prediction from the aerial platforms above urban environments. In [30], the authors introduced the drones' coexistence techniques and distributed IoT devices on the ground-based stations in Machine to Machine (M2M) communication networks. Emerging advanced technologies such as 4G/5G/6G have potential applications for UAVs equipped with IoT devices and Global Positioning System (GPS) receivers for delivering IoT services in many applications from great heights. The authors of [73] discussed the importance of UAVs with advanced intelligence techniques to cope with the delay and minimize overhead limitations of Vehicle Ad hoc NETworks (VANETs).

Many scenarios need a UAV to gather data from different real-time applications in the smart world, such as public safety, dynamic agriculture coordination, disaster management, recovery, data routing, monitoring accident, intelligent transport systems, intelligent flying BS [74], and relay data from isolated sensors to a BS. Table 2 illustrates the scenarios in which UAVs have already been used to gather data for various applications. The UAVs that carry IoT devices can also store and process data to enable a UAV to perform tasks effectively. Due to the energy required for processing IoT tasks and data urgency, the authors of [75] suggested that data processing should be performed locally. After that, the processed data should be forwarded to the server for taking necessary actions. Data gathered by UAVs has been collected to assist the research teams in serving so many applications.

Table.2 Scenarios where UAVs are used for data collection

| Data collection | Scenarios | Ref. | Ideology | Advantages UAV in the scenario |
|---|---|---|---|---|
| **General** | Dynamic coordination and data routing | [76] (2017) | • The UAV can be used as a relying station to maintain connectivity of the UAV and IoT devices. | ➢ Building a UAV backbone network and keeping it connected at all times.<br>➢ Real-time communication is essential; the UAV will move inside the transmission range and maintain its connectivity.<br>➢ In this scenario, real-time communication is not essential; UAVs may leave the communication range, gather data, store data, and come back to the required dominator's communication range and data. |
| | Monitoring accidents | [77] (2018) [4] (2020) [78] (2012) | • UAVs as an integral solution to aid the SAR in the accident scene within the shortest time.<br>• UAVs provide the SAR an advance report before reaching the incident scene. | ➢ The UAV can rapidly fly over high-traffic streets and reach the accident location quickly.<br>➢ The UAVs reach the accident location in real-time and can send a detailed report of the situation.<br>➢ The UAV can establish a real-time communication channel between SAR still on their way and the accident site. |
| | Intelligent Transportation Systems | [76] (2017) [79] (2019) | • The reliable and efficient transportation system. | ➢ Improving traffic<br>➢ Bringing better safety and improving security on the road<br>➢ Enhancing driver comfort |
| **UAVs collect data from IoT devices on the ground** | Data relaying | [80] (2017 | Links between the sensors on the ground and BS. | ➢ Maintaining the connectivity |
| | Disaster management | [81] (2018) [82] (2019) [83] (2021) | • Disaster management by using big data gathered by IoT.<br>• Providing an agile survey of the disaster area to speed up the SAR operation. | ➢ The UAV is used as $0^{th}$ responders for delivering communication service to the victims.<br>➢ Data is gathered and analyzed locally to search the optimal position. |
| | Energy efficiency | [84] (2016) [85] (2017) [86] (2019) [1] (2016) | • Optimal flying trajectory and deployment swarm UAVs to collect data from ground IoT devices.<br>• Maintain energy-efficiency in time-varying mobile IoT networks.<br>• Swarm UAVs collect data from IoT devices on the ground. | ➢ Data gathering.<br>➢ Energy-efficiency. |
| | Drone- wireless sensors networks | [87] (2017) | • Data collection from sensors nodes<br>• Large coverage area. | ➢ Data gathering.<br>➢ Delivery service to a large area. |
| **UAVs are equipped with IoT devices** | A UAV equipped with an IoT device to perform specific tasks | [88] (2018) [82] (2019) | • IoT sensors collect data in real-time, which can be processed on board in case enough power is available or transmitted to BS. | ➢ The UAV may be equipped with radars to provide resolutions from a submillimeter to a few centimeters and high-end electro-optical sensors. |
| | | | 1. Monitoring and Surveillance applications. | ➢ The UAVs are equipped IoT devices at the lower end of the spectrum to map, locate, and sense obstacles. |
| | | | 2. Monitor environmental and weather conditions. | ➢ The UAVs carried thermal sensors to monitor ecological and climate conditions. |
| **UAVs are equipped with IoT devices and collaborated with IoT devices** | Keeping connectivity | [89] (2021) [90] (2021) | • Single strength prediction is suitable for all of the above applications and any real-time applications. | ➢ Maintaining connectivity.<br>➢ Provisioning optimal QoS.<br>➢ Identifying UAV coverage area. |

On the other hand, massive IoT devices process and store data on the devices themselves, but they involve large hardware and power requirements. Therefore, the drone's use will reduce energy usage and control of both sleep and active IoT devices for greening IoT. The authors of [91] consider a UAV as a sink node in a wireless communication network to collect data from the sensors at any event. They also investigated the optimization of the UAV network service area. Processing data on the UAVs is required for data-intensive operations such as dynamic sense, image processing, and path planning. In [92], the authors designed efficient data processing devices to process data in low power consumption. Furthermore, in [93], the authors employed Machine Learning (ML) techniques to an image captured by IoT devices such as cameras onboard a drone. A Convolutional Neural Network (CNN) with a Support Vector Machine (SVM) was used for the image recognition tasks.

Briefly, gathering data by using UAVs can be done in several ways: (i) UAVs collecting data from IoT devices deployed in any environment and then delivering gathered data to the nearest BS [94], (ii) UAVs equipped with smart IoT devices to collect data directly from any problematic place to reach [95], (iii) UAVs equipped with smart IoT devices and collaborate with smart IoT devices in smart cities. For example, UAVs are utilized for public safety and data collection from SAR's smart wearable devices to perform specific tasks.

### III. Energy Efficiency

The UAV can provide energy efficiency for IoT devices by minimizing their power consumption. To save energy while sending the data captured by the IoT devices, the UAVs can move toward IoT devices to gather data. The UAVs advanced communication technologies such as B5G can exchange data with IoT devices in smart environments. The authors in [16] applied the genetic algorithm to improve drone-assisted IoT sensor networks based on energy consumption, sensor density, flight time, and fly risk level. The authors of [96] summarized the strategies for green cellular IoT based on energy saving.

Applying controlling and processing techniques in UAVs are the essential points in saving energy. A UAV is equipped with IoT devices to process, manage the data, and then deliver it efficiently, taking advantage of LoS with the transmitters [75]. A cooperative framework for the drone-based wireless sensor network (WSN) was presented in [97]. It composes of fixed-group leaders, drone-sink, and sensor nodes. This study showed that the energy consumption and processing complexity of the leader's election are reduced. In [98], the authors discussed the strategies and techniques of drone-based WSN for collecting data. The proposed techniques and strategies could reduce energy consumption, extend flying time, and reduce data collection latency. In [99], the authors presented an algorithm to collect data in WSNs by applying UAVs and mobile agent technologies. Both techniques were shown to contribute to saving time and energy consumption of sensor nodes. More so, in [100], the authors proposed a model of one IoT device's efficient energy.

Additionally, the authors of [101] explored the importance of drones' cooperation with WSNs to improve energy-efficiency by relaying to extend their lifetime. The outcomes showed that the proposed technique reduced delay and improved UAV routing with a better coverage area. In [102], the authors proposed an efficient power consumption model to consider the load factor and speed for the drone-based relay. Furthermore, a wired UAV docking system was presented to implement several functions via the collaboration of UAV and IoT devices to improve energy efficiency, reduce wasted resources, and ensure security among them [103]. In [104], the authors introduced drones' applications for improving IoT security platforms, emergency response, and building monitoring by utilizing beacons. Research targeted towards extending the battery life in UAVs has also been performed. Authors of [105] discussed an automatic battery as a replacement technique, while an automatic battery allowed the UAV to keep working indefinitely without any manual intervention. This mechanism successfully looked into the power management in UAVs for continuous surveillance indoor and outdoor environments.

For UAV- deployment energy efficiency, the authors of [102] introduced an energy efficiency approach using a channel model. The communication of UAVs with energy efficiency was investigated with trajectory optimization [106]. The authors of [107] introduced the use of UAVs to enable MECs and measure the optimal altitude and throughput. Furthermore, designing UAVs communication trajectories improve energy efficiency [56, 108]. The trajectory optimization uses for recharging UAVs using Q-learning [54] and balances energy consumption and QoS [53]. In [109], the maximization of UAVs' energy efficiency is discussed, in which the UAVs are deployed as a relay station to amplify the signal strength between the IoT devices and end node (center). The authors of [110] focused on the energy consumption of UAVs and optimal MECs offloading. The energy harvesting IoT performance by using UAVs with Nakagami-m fading is discussed in [111]. Table 3 summarizes the advantages of using UAVs for enhancing the energy efficiency towards greening IoT.

Table.3 UAV for energy efficiency to green IoT

| Ref | Highlight | Advantages | Limitations and Suggestions for Improvement |
|---|---|---|---|
| **[16](2016)** | Drone-assisted IoT sensor networks. | ✓ Improving energy consumption.<br>✓ Increasing flight time and reducing the level of fly risk. | ✓ Battery lifetime.<br>✓ Data gathering and processing. |
| **[98](2017)** | Drone-based WSN to gather data. | ✓ Reducing energy consumption<br>✓ Reducing latency and flying time during data collection. | ✓ UAVs collaboration for efficient data acquisition in a large coverage area. |
| **[99](2014)** | Gathering data by employing UAV and WSN. | ✓ Saving time with the energy efficiency of sensor nodes. | ✓ Parallel processing for multi-UAV coverage area for saving time and response. |
| **[101](2016)** | Cooperation of WSN and UAV technology. | ✓ Energy-efficiency of UAV relaying. | ✓ Clustering techniques can improve energy efficiency and collaboration of WSN. |
| **[112](2016)** | UAV docking system and IoT collaboration. | ✓ Reducing sources of waste and energy by ensuring the security. | ✓ UAV coverage area. |
| **[84](2016)** | Optimal flying trajectory of swarm UAVs to gather data from IoT devices. Maintaining energy-efficient communications in the time-varying mobile IoT networks. | ✓ Data gathering<br>✓ Energy-efficiency | ✓ Coordination of multi-UAV for large coverage area. |
| **[113](2017)** | Tradeoff between the number of update times, the mobility of the drones, and IoT devices transmit power. | ✓ Energy-efficiency<br>✓ Improving QoS | ✓ UAV efficient trajectory for data gathering.<br>✓ Optimal position of UAV. |
| **[114](2019)** | Importance of edge computing for IoT environment in smart energy. | ✓ Reducing energy consumption.<br>✓ Reducing data transfer from the IoT to edge and then to the cloud.<br>✓ Reducing computing energy and the cost of network resource. | ✓ Data latency |
| **[115](2019)** | Development of the resource allocation of UAV for improving energy efficiency. | ✓ Maximizing energy efficiency. | ✓ Designing trajectory and link of UAV wireless networks |
| **[57](2020)** | Reactive RF band Allocation for improving QoS and energy efficiency. | ✓ Trade-off between energy consumption and QoS. | ✓ Converge is not taken into consideration |
| **[108](2020)** | Designing trajectory for UAV communication with improving energy efficiency. | ✓ UAV speed and trajectory.<br>✓ Maximizing throughput.<br>✓ Energy efficiency. | ✓ Coverage area |
| **[96](2020)** | Strategies for energy efficiency of cellular IoT. | ✓ Saving energy. | ✓ Massive devices connectivity and charging policy. |
| **[109](2020)** | Improving the energy efficiency of UAVs-enabled IoT. | ✓ Improving energy efficiency.<br>✓ Achieving green communication. | ✓ Fixed UAV trajectory. |
| **[110](2020)** | Minimum energy consumption of UAV with optimal MEC offloading. | ✓ Minimizing UAV energy consumption.<br>✓ Optimal trajectory. | ✓ Uncertainty of user mobility. |
| **[116](2020)** | Energy harvesting and energy efficiency for energy shortage. | ✓ Improving energy efficiency. | ✓ Achieving a green radio environment. |
| **[55](2020)** | UAV for data gathering. | ✓ Optimal trajectory.<br>✓ Enhancing QoE.<br>✓ Energy efficiency. | ✓ Deep neural network is reqired to predict Q-values in the large Q table. |
| **[111](2021)** | IoT energy harvesting using UAV. | ✓ Minimizing outage probabilities.<br>✓ Multi-antenna UAV-enabled relay (UR). | Multi-hope UAV-enabled relay to enhance the outage probabilities and throughput for the applications of IoT. |

## IV. ICT technologies

Green information and communication technology (ICT) plays a vital role in green IoT devices. Green ICT focuses on green RFID technology, green sensor networks technology, green cellular networks, etc. [117]. IoT poses a high potential to bolster environmental sustainability and economic [105] with the advancements of enabling technologies. The evaluation and impact of ICT on power consumption and $CO_2$ emissions, which are increasing rapidly and damaging our environment, are highlighted in [118].

### 1) RFID

Reducing RFID tag size and developing energy-efficient algorithms to optimize the tag estimation are needed to green RFID [119, 120]. Hubbard *et al.* [121] addressed the techniques required to enhance UAVs' battery lifetime and the importance of RFID reader detection range. The benefit of integrated RFIDs and UAVs is the support of additional information that could be used in the supply chain of management systems. In the same context, in [122], the authors used a feasibility analysis to recharge the multipurpose RFID tags in UAVs for monitoring environment with energy balancing demonstration.

Furthermore, the authors of [123] presented an efficient solution to monitor harsh environments using the combination of UAVs and RFIDs. The finding showed that the UAV equipped a set of RFID tags and sensors to monitor a specific area. Furthermore, the findings showed that the combined technologies worked correctly, and tags had the power to monitor instruments, particularly where the monitoring is required for a harsh environment. UAVs are used to gather data from RFID and scattered throughout the area [123].

The combinations of RFID and UAV work in conjunction, and tags are powerful for monitoring the instruments while the monitoring is required to deliver service in a large area in harsh environment. Moreover, in [124], the authors introduced UAVs with indoor localization techniques using RFID systems. Their main aim is to achieve simplicity and cost-efficiency. UAV tracking and localization are considered to achieve simplicity and cost efficiency by using radio-frequency shadowing, observation of received signal strength indicator (RSSI), and interference under indoor environments. A summary of using a UAV with RFID for green IoT is shown in Table.4.

### 2) Wireless Sensor Networks

Sensors are deployed around smart cities for data gathering and measuring global and local environmental conditions. A sensor node contains several small, low-cost, and low-power electronic devices such as a battery, sensing unit, a storage unit, a processing unit, and a communication unit [125]. These nodes suffer from scarce resources (i.e., limited sensing range, battery power, communication range, storage, process, and computing resources) [126]. Every sensor can read data from its surroundings such as sound, temperature, pressure, acceleration, humidity, etc. [127]. Furthermore, sensors collaborate and communicate with each other in a sensing environment based on 5G and deliver the BS's required sensory data using ad-hoc technology. Mehmood *et al.* [128] discussed the smart control of energy-efficiency in routing protocols for WSN concerning the design trade-offs. Traditional routing protocols in WSNs are generally challenging to maintain desirable network performance due to the unstable environment situations and application requirements. The sink node will drive some relay nodes in the path to deplete their batteries earlier than others. Unbalanced-energy consumption routing results in a shorter network lifetime. In contrast, dynamic path routing with the assistance of probabilistic forwarding is a proper solution since it distributes the network's traffic load and the energy consumption in the network. Such challenges are discussed in [129] [127].

WSNs have been used for various applications such as environment monitoring [125, 130-132], fire detection [133-135], object tracking [136-138], evolving constraints in the military [139], monitoring machine health, and monitoring the industrial processing [125]. Green IoT benefits from keeping the sensors mode in sleeping to save energy [140, 141]. Recently, sensor nodes gain energy harvested from their environment, including sun, kinetic energy, vibrations, temperature differentials and so on [142-144]. WSNs supply sufficient power to improve the lifetime of the system and reliable transmission. IoT networks require a large number of sensors deployed for monitoring a specific environment.

To enhance UAVs function, Malaver *et al.* [145] discussed integrating UAVs technology and WSN technology. The $CO_2$ concentration was monitored during the data collection in real-time by integrating UAV and WSN technologies. The integrated system is recommended and applicable for many applications, including mining studies, agriculture, bushfires, botanical and zoological domains [146]. Besides, the authors of [147] introduced an efficient technique called Particle Swarm Optimization (PSO) for analyzing data gathered by integrated UAV and WSN technologies systems. Further, integrated UAV and WSN technologies are used for data gathering with low power from large coverage areas [148]. The integrated system indicated an efficient solution for energy utilized resources [97, 101]. The UAVs can manage node duty cycling status, i.e., sleep or active.

Furthermore, WSNs and smart drones' cooperation was utilized to reduce energy consumption during the routing process, which highly maximized IoT devices [101]. Moreover, the authors of [149] presented the communication links between the WSN nodes and UAVs to optimize wireless medium scheduling and UAV trajectory setting for reducing nodes' energy consumption while gathering the sensory data from the target environment. This idea is supported by a study in [150], which revealed that energy consumption is reduced due to a drone's use, which can move sequentially closer to the target environment and gather WSNs nodes' data from short distances. Moreover,

Rashed et al. [87] presented the tradeoff between operation time and nodes' coverage to choose the right mobility pattern for UAVs. Data takes two ways to gather data, i.e., data transmitted from the sensor nodes on the WSN environment to the UAV and data gathered by sensor nodes equipped in the UAV board. Collected data is sent from the UAV to the cloud for analysis. Furthermore, wireless medium allocation and energy allocation of sensory data in the first and second ways are considered and discussed deeply [151]. Table. 4 shows a summary of methods for using a UAV with sensors for green IoT. The combination of UAV and WSN for monitoring extensive scale application is summarized in [152]. The collaboration between the WSN and UAVs becomes the cheapest and friendly for monitoring smart applications in the real world with intelligence techniques. The authors of [153] introduced the collaboration between the UAV, WSNs, and IoT for crop monitoring and precision agriculture.

Briefly, with the help of UAV, the communication bandwidth between the UAV and WSN is enhanced, the energy required for data transmission is reduced, and the coverage area is extended. Therefore, the monitoring and data gathering from WSN improve efficiently. The collaboration of WSN and UAVs plays a vital role in improving energy efficiency, processing locally, making decisions, and extending the coverage using clustering techniques, as shown in Fig.5.

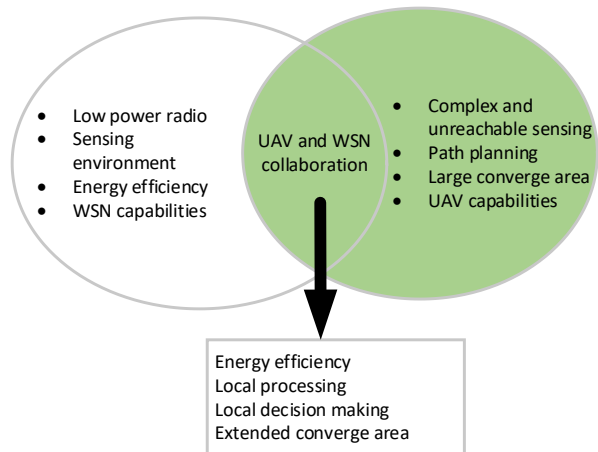

Fig.5 WSN and UAV collaboration

### 3) B5G

Nowadays, 5G represents a promising technology that impacts our life and environment, even in improving the IoT performance, as shown in Fig. 6. The B5G applications include robotic communications, e-health, the interaction between the robots and humans, media, e-learning, public safety, e-governance, transport and logistics, industrial and automotive systems. Green wireless communication technologies play a crucial role in greening IoT. It refers to sustainable, energy-efficient, energy-aware, and environmentally-aware technologies. The green communication network idea is aimed at reducing $CO_2$ emissions, radiation exposure, and energy consumption. Abrol et al. [154] presented the growing and influence need for energy efficiency. Energy efficiency adaption fulfills the demands for enhancing data rate, increasing the capacity, and providing high QoS of the next-generation networks (NGN). Many research types have been done to save energy by using solar power to enhance QoS [50, 155-160]. Koutitas et al. [161] used genetic algorithm optimization to develop network planning. The finding showed significantly $CO_2$ reduction, cost-saving, and low radiation exposure. In the same direction, Naeem et al. [162] disused how to minimize $CO_2$ emissions and maximize the data rate in the cognitive WSNs environment.

The study in [163] discussed a set of system parameters in order to evaluate the energy consumption and the $CO_2$ emissions of wireless communication networks. Moreover, Motlagh *et al.* [164] presented a steering technique among multiple 4G networks for ensuring efficient and reliable network connection from drones, in which the UAV improves the QoS and costs to an acceptable amount of energy. Also, Xu [165] proposed a single UAV framework representing a sink node and a head cluster of sensors in a specific environment, considering a low message delay. On the other hand, the broad UAVs' energy is a challenge, while protocols should help the UAVs network [166].

Therefore, UAVs are a promising technology to delegate wireless services to large coverage areas with reducing energy consumption. The authors of [167] discussed the energy efficiency of heterogeneous 5G cellular networks with the help of UAVs' help. They formulated how the efficient coverage of UAVs is improved with energy efficiency. A summary of the advantages of using UAVs with 5G technology for green IoT is shown in Table 4. The development of an integrated UAV and B5G network with intelligent techniques can consider balancing the extended coverage area at the cost of energy consumption or increasing the number of UAV with improving QoS.

Table.4 UAV and ICT technology for green IoT

| Ref | Technology | Highlight | Advantages | Limitation |
|---|---|---|---|---|
| [168] (2016) | RFID | Enhancing the lifetime of UAV battery. | Providing additional information. | RFID implementation and integration for the management process of supply chain. |
| [169] (2015) | RFID | Monitoring operation in harsh environments. | Powerful monitoring instruments and large coverage. | Monitoring different activities. |
| [170] (2012) | RFID | UAV indoor localization. | Cost efficiency and simplicity of UAV localization and tracking. | Outdoor tracking and localization. |
| [148] (2017) | Sensors | An Integrated UAV-WSN system for data collection. | Low-power data gathering from a large coverage area. | Monitoring different applications for giving time interval. |
| [151] (2014) | Sensors | Transmission data. | Bandwidth enhancement. Reducing energy consumption. | Interference among multi-UAVs during data collection of the same environment. |
| [150] (2016) | Sensors | Optimal UAV trajectory for data gathered from IoT. | Reducing transmission energy. | Optimal coverage probability. |
| [145] (2015) | Sensors | Integration of UAV and WSN. | Monitoring $CO_2$ emission. | Advanced technologies such as solar cells, batteries, and sensing for data acquisition. |
| [149] (2017) | Sensors | Communication links between the WSN environment and UAVs. | Optimizing scheduling of UAVs trajectory for minimizing energy consumption of WSN. | Optimal trajectory and coverage area. |
| [161] (2010) | B5G | Network planning development. | Reducing $CO_2$ emission, low cost and low exposure to radiation. | Green deployment of B5G networks. |
| [164] (2017) | B5G | UAV connectivity among multiple 4G networks. | High QoS. | Balancing QoS and energy consumption. |
| [165] (2016) | B5G | UAV as a mobile sink. | Reducing delays. | Optimal placement techniques for the ground node and UAV. |
| [171] (2020) | 5G | UAV-based 5G design using ML. | Improving QoS. | Multi antennas using with ML. |
| [172] (2020) | 5G | Resource allocation and energy harvesting for green IoT. | Enhancing energy efficiency. | QoS and QoE. |
| [173] (2021) | 5G | UAV trajectory planing in 5G networks. | Improving throughput. Maximizing the flight distance. Improving QoS. | Trade-off of energy efficiency and QoS. |
| [153] (2020) | WSN and IoT | Collaboration of UAV, WSN, and IoT for crop monitoring. | Efficient data collection. Precision agriculture performance. | Coverage area and energy efficiency and computing of collaboration of UAV, WSN, and IoT. |
| [23] (2021) | 5G | UAV and data aggregation for enabling IoT. | Improving sinking ratio. Reducing energy consumption. | QoS cost and coverage area. |

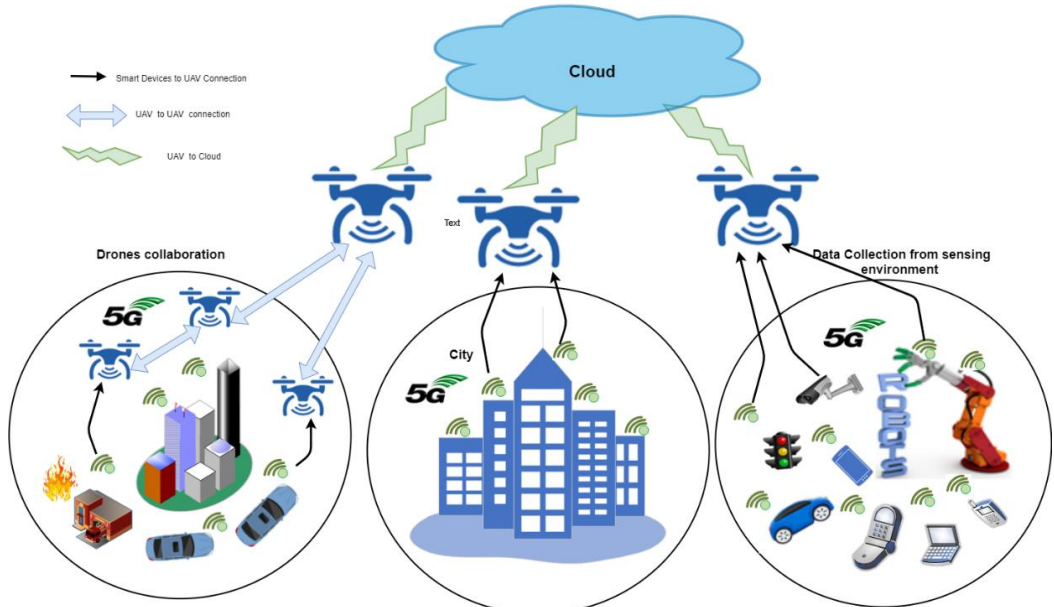

Fig.6 UAV expected 5G for greening IoT

## V. Hazardous Pollution

Newly monitoring environmental pollution is an essential concern in our world. WSNs can be utilized to monitor environmental pollutions. However, the transmission power of sensor nodes is limited to transmit data in real-time.

Thus, the authors of [174] developed a UAV system capable of measuring point source emissions. The study focuses on the airflow behavior and the performance of CO, $CO_2$, $NO_2$, and NO sensors for measuring the emissions in a particular area. The greenhouses' pollution should also be taken into consideration for controlling the gas emission from the greenhouses. The study in [146] emphasized the importance of a solar-powered UAV equipped with a CO2 sensing system integrated with a WSN to control pollution produced from the greenhouses. Flexible deployment has been found an appropriate technology for air pollution monitoring [175]. Authors in [176] studied the existing UAV techniques for monitoring the application environment. Heuristic and PSO techniques monitor the particular area and focus on the most polluted zones [177].

Moreover, the study in [178] introduced an efficient module for UAV platform to monitor multiple air pollutants in real-time. The study in [179] improved the idea of combining UAVs and mobile stations, which can be navigated autonomously for pollution monitoring. The UAV was used as a platform in air quality research, focusing on the roadside air pollution profiling. However, Zang et al. [180] demonstrated experiences in applying UAVs in the water pollution investigation in Southwest China in the presence of high altitude, low air pressure, strong air turbulence, severe weather, and cloud cover. The authors of [181] summarized the UAVs' importance for monitoring hazardous pollution and emission of chemical materials. Simultaneously, the authors of [182] focused on designing UAVs for monitoring the air quality base on IoT technology. The finding shows that air pollution data was efficient and in real-time.

## VI. Use-Cases

UAVs are the merger of various mature and advanced technologies that have rightly attracted much attention worldwide. Due to their appeals for smart applications, their deployment is a new potential market. The primary objective of developing a drone-based system is to meet large operational areas, high mobility, low cost, rapid deployment, and acceptable positioning accuracy. Connecting the UAVs will enable many new services and business opportunities. Klimkowska et al. [175] discussed using UAVs for various maritime applications such as red tide detection, ship detection, monitoring, border patrol tracking, and the ability to fly over hazardous and dangerous areas.

Green IoT makes smart devices connect to the real-world efficiently. The numerous green IoT applications include smart homes, smart cities, smart grids, smart agriculture, and smart healthcare. The data-gathering potential of UAVs and IoT devices is enormous. The drone's ability to fly over crops, houses, and cities makes a perfect platform solution to collect data that was once too expensive to gather and difficult for a human to do. It is envisaged that the things around us will be smarter and greener because of a drone's capability to fly rapidly and collect data in real-time.

### A. Smart Homes

Recently, UAVs have been creating enormous development in our society, bringing new leaps toward home surveillance and security. UAVs integrated with IoT devices like cameras are envisaged for security and surveillance of smart homes, company establishments, and government offices. For secure smart home applications, a tethered UAV can carry the required monitoring devices in the payload. It uses solar-powered sensor nodes for sensing physical movements, as well as any acoustic signals. For such applications, the UAV has to remain in the proximity of the house/office under surveillance. The wireless technique is also used for charging the battery of UAVs and the IoT devices in the UAV payload.

### B. Smart Cities

Recently, UAVs are being used to improve the life quality in smart cities by documenting accident and emergency scenes, providing first responder activities in disaster and emergency cases, and monitoring construction sites [183-185]. Smart cities need smart sensory data. IoT devices are used to collect data, while UAVs are used as edge computing nodes to collect and process the data from IoT devices. Integration of UAV and IoT plays a critical role in enhancing the quality of life for city residents. Furthermore, decreasing the number of IoT devices can be done using a drone, which can be used to save energy and reduce the $CO_2$ emission and pollution of IoT devices. Won et al. [186] presented a study on secure communication protocols between the UAV and IoT devices for refining the elegance of smart cities. The protocols deliberated include multiple kinds of fixed IoT devices and mobile devices with different capacities. Numerous UAVs in forthcoming smart cities viz. privacy, cybersecurity, and public safety are discussed in detail [187]. Furthermore, UAVs applications for smart cities and the opportunities to crack today's problems have been discussed in [186, 188]. Lately, crimes, such as terrorism, vandalism, and street crimes, have increased in urban areas. Therefore, anticipating crimes through the detection of criminals among people crowds is vital. In this vein,

UAVs can be used to detect, track, and recognize criminals, follow their movements, thereby providing immunity from any unwanted situation.

UAVs have also been considered as relay stations in other research directions to collect and transmit data from IoT devices in many applications such as emergency services, healthcare, pollution monitoring, traffic tracking, surveillance, fire detection, earthquakes, etc. With the use of drones, deployment time, cost, and energy consumption are reduced. Consequently, UAVs are graceful to become an integrated part of the smart cities communication network for being deployed to quickly collect, analyze, aggregate, and deliver highly accurate and detailed data.

### C. Smart Agriculture

Agricultural UAVs are a modern technique used in farming to be used for aerial imagery with significant efficiency. UAVs offer useful data processing capabilities via cloud-based computing in delivering aerial monitoring applications in real-time and intelligence-gathering. The study in [189] explained several potential applications that required the UAVs to collaborate with sensors to perform tasks efficiently, such as remote crop monitoring, soil moisture sensing, infrastructure monitoring, monitoring water quality, and remote sensor deployment. UAV fogs are required to replace massive IoT devices, which will decrease pollution and power consumption. UAV's low cost and efficiency are essential in agricultural operations for monitoring and helps farmers know all things about their fields [190]. In [191], Almalki proposed an adaptive drone model to detect food quality and safety. Briefly, UAVs are used in agriculture for providing transformation, data collecting, and monitoring fields regularly.

### D. Smart Grid

The UAV-IoT equipment maintains power distribution lines, and natural gas pipelines [192]. Nowadays, one of the most significant drone's applications is to carry out the often dangerous work of power crews in investigating storm damage or maintaining power lines. Hence, UAVs integrated with IoT devices can find and inspect downed power lines quicker than manned crew members. Also, UAVs can cover larger areas while protecting crew members from danger. UAVs can transmit real-time data and live video streams toward the smart grid systems in such an application, allowing power consumption to be managed by potential electrical issues like never before. Consequently, UAVs decrease both inspection times and costs [193]. In [194], discussed the potential solution for communication and management of smart grid domain with the help of edge UAV to avoid the delay-tolerant network of UAV and IoT network.

### E. Smart Healthcare

UAVs play a vital role in picking and fast delivering vaccines, medicine, and blood samples to any places. Currently, UAV is used to comb COVID-19 based on the above functions and monitor and avoid human interaction [7, 195]. Furthermore, it enables delivering efficient healthcare to patients, including blood, birth control pills, medical supplies to rural areas, vaccines, and snakebite serum [196]. Reaching on time, in some cases, could mean the difference between death and life. Therefore, the UAV can reach victims who need immediate medical attention within minutes. UAVs also can transport medicine within the hospital as well as carry blood between hospital buildings. Furthermore, UAVs give elderly patients tools to support them and offer various exciting possibilities to the healthcare industry for saving time, cost, and lives. UAVs can also deliver medical supplies and food aid to areas hit by natural or human-made disasters. Therefore, drones, along with wearable and implantable technologies, will revolutionize healthcare due to their ability to continuously work alongside IoT sensors to monitor people's health continuously. By utilizing these combined technologies, patients may not need to travel miles to get a diagnosis.

## VII. Future scopes and insights

Tremendous efforts have been made in making IoT technologies greener. However, they still face significant challenges from many aspects, such as limited energy, pollution hazards, and QoS. UAV technology is finding its way into greening IoT, and it is considered the future of IoT. Despite the tremendous advantages of the UAV edge intelligence concept for delivering service in real-time and providing high-reliability services to IoT devices in several applications, reducing energy consumption, reducing delays, and providing efficient bandwidth, there are still several open issues and challenges as discussed in the following. The authors addressed the convergence of UAV edge intelligence and blockchain for solving the challenges in B5G networks for Industry 4.0 applications [197].

### A. Efficient UAV trajectory

Adjusting the trajectory of the UAVs is challenging. Therefore, designing an efficient trajectory to meet computing tasks is another open research problem. While considerable research has been dedicated to UAV's trajectory planning, developing autonomous and efficient trajectory planning techniques taking into the account various factors such as task priority, delay, user locations, and energy consumption is still an open research direction.

## B. Energy efficiency

The limited battery capacity of the UAVs represents a critical constraint. Apart from the drone's energy consumption and communication, a UAV also needs energy for computational tasks. The field of energy-efficient propulsion of drones, in which the required energy on a UAV for computing and communication is optimized, still needs further investigation. Furthermore, recharge scheduling and energy harvesting may be considered as another direction for improving the drone's battery capacity. Furthermore, clustering users or IoT devices to collaborate for saving energy is another viable direction of research. For example, each user/ vehicle/ IoT device can serve as the neighbor inside the cluster to reduce energy consumption. Since the distance of transmission between neighbors is shorter than that of the drones, the energy efficiency may increase significantly in such cases. Regarding the reviewed current literature focusing on UAV edge intelligence for greening IoT in different applications and from different aspects such as energy efficiency, connectivity, and pollution hazardous, we believe more future research is required to be done as shown in Table 5. Furthermore, we addressed the challenges of UAVs during performing tasks in order to green IoT, as shown in Table 6.

Table.5 Research directions, challenges, and insights

| Study | Directions | Challenges | Future insights |
|---|---|---|---|
| [35] (2016) | Drone-based IoT services | Managing a massive number of UAVs for such reasons: UAVs equipped with different IoT devices Different videos are taken from different angles of UAV position | • It is essential to find an efficient scheme to manage and control IoT devices' power conserve in the UAV payload. • Security is a critical issue on the drone-based IoT services, and therefore, finding accurate and efficient techniques to avoid hacking UAV communication is required. |
| [198] (2014) | Flight time enhancement | Battery lifetime | • Battery lifetime needs improvement to allow UAV flying over long-distance and increase the time of UAV flight. |
| [92] (2017) | Real-time application Real-time processing Accurate decision making | Collecting a large amount of unlabeled data becomes less expensive solved real-world issues | • Deep learning technique is required to be implemented in the payload of the UAV to demonstrate data analysis by low power and efficient DL in support of real-time applications |
| [199] (2016) | Energy consumption | Energy Limitations | • Finding suitable techniques for efficient batteries, new lighter materials, and energy harvesting may lead to UAV technology's potential applications. |
| [93] (2017) | Intelligent techniques | UAVs limited battery Real-time and reliable communications with the BS given QoS and energy constraints | • Implementation and design of an efficient power distributed algorithm is required for real-time processing of swarm drones' images, captured videos, and gathering data. |

Table.6 Summary of UAV challenges and proposed solutions for greening IoT applications

| Issue | Challenges | Solutions |
|---|---|---|
| Energy consumption | UAV battery | ✓ Optimal trajectory 3D space ✓ Solar panel ✓ Drone tethered for fixed-function ✓ Energy harvesting techniques |
| Connectivity | Increase number of users Or IoT devices | ✓ Designing a drone with more efficient network to solve the scalability issue (e.g., using blockchain technology with sharding techniques). ✓ UAVs-based aerial caching |
| Controlling | Collision | ✓ Blockchain technology for decentralized multi-UAV autonomously |

## VIII. Conclusion

IoT improves and simplifies our lives in numerous ways to protect our environment, and society by sensing and cooperatively communicating over the internet. Finding suitable and efficient techniques for greening IoT is required to improve our quality and sustainability of resources. Collaborative UAVs and IoT have the potential to change our daily lives drastically and make a smarter and greener world. UAVs equipped with IoT devices are used in many applications to perform complex tasks efficiently at a low cost. We found that UAVs technology plays a critical role in reducing energy, reducing pollution, and enhancing the connectivity and QoS of IoT devices. Throughout this survey, we discuss the importance of emerging UAVs for greening IoT based on efficient techniques that can enable green ICT technologies, increase energy efficiency, and tackle pollution hazards. We thoroughly identify key challenges and future insights of UAVs for green IoT. We hope that the critical research challenges and research insights described in this paper will help specify the way for researchers to improve collaborative UAVs for green IoT applications in the future.

**Funding**
This project has received funding from the European Union's Horizon 2020 research and innovation programme under the Marie Skłodowska-Curie grant agreement No. 847577; and a research grant from Science Foundation Ireland (SFI) under Grant Number 16/RC/3918 (Ireland's European Structural and Investment Funds Programmes and the European Regional Development Fund 2014-2020).